\documentclass[12pt]{article}
\usepackage{epsf}
\usepackage{psfig}
\usepackage{epsfig}
\usepackage{graphicx}
\usepackage{amssymb}

\def\di{\displaystyle}

\def\eq#1{(\ref{#1})}
\def\Eq#1{Eq.~(\ref{#1})}

\def\Tr{{\rm Tr}}

\def\s0#1#2{\mbox{\small{$ \frac{#1}{#2} $}}}
\def\0#1#2{\frac{#1}{#2}}

\def\eq#1{(\ref{#1})}
\def\Eq#1{Eq.~(\ref{#1})}

\def\di{\displaystyle}

\def\bg{\begin{eqnarray}\begin{array}{rcl}\displaystyle}
\def\eg{\end{array} &\di    &\di   \end{eqnarray}}
\def\bm#1{\begin{eqnarray}\begin{array}{#1}\di}
\def\bmo#1{\begin{eqnarray*}\begin{array}{#1}\di}
\def\bml#1#2{\begin{eqnarray}\begin{array}{#1}\label{#2}\di}
\def\bgo{\begin{eqnarray*}\begin{array}{rcl}\displaystyle}
\def\ego{\end{array} &\di    &\di \nonumber  \end{eqnarray*}}

\def\btensor#1#2{\renew\left#1\begin{array}{#2}\di}
\def\brtensor#1#2#3{\ren#3\left#1\begin{array}{#2}}
\def\botensor#1#2{\renew\left#1\begin{array}{#2}}
\def\etensor#1{\end{array}\right#1}

\def\eq#1{(\ref{#1})}
\def\Eq#1{Eq.~(\ref{#1})}

\def\Tr{{\rm Tr}}

\def\T{{\rm T}}
\def\id{1\!\mbox{l}}

\def\be{\begin{equation}}
\def\ee{\end{equation}}
\def\bea{\begin{eqnarray}}
\def\eea{\end{eqnarray}}

\def\h{\hbox{$\frac{1}{2}$}}

\font\af=msbm12
\def\T{{\mbox{\af T}}}
\def\R{{\mbox{\af R}}}

\date{\today}

\def\ren#1{\renewcommand{\arraystretch}{#1}}

\def\renew{\renewcommand{\arraystretch}{1}}
\ren{1.5}

\begin{document}

\begin{flushright}
INLO-PUB-02/03\\
FAU-TP3-03-02\\
\end{flushright}
\par
\vskip .5 truecm
\large \centerline{\bf Doubly Periodic Instantons and their Constituents}
\par
\vskip 0.5 truecm
\normalsize
\begin{center}
{\bf C.~Ford} ${}^{a}$ and
{\bf  J.~M.~Pawlowski} ${}^{b}$
\\
\vskip 0.5 truecm
${}^a$\it{
Instituut-Lorentz for Theoretical Physics\\
Niels Bohrweg 2, 2300 RA Leiden, The Netherlands\\ 
{\small\sf ford@lorentz.leidenuniv.nl}\\
}
                                            
\vskip 0.5 truecm
${}^b$\it{Institut f\"ur Theoretische Physik III,
Universit\"at Erlangen\\
Staudtstra\ss e 7,
D-91058 Erlangen, Germany \\
{\small\sf jmp@theorie3.physik.uni-erlangen.de}
}

\vskip 0.5 true cm

\end{center}

\vskip .7 truecm\normalsize
\begin{abstract} 

Using the Nahm transform we investigate doubly periodic
charge one $SU(2)$ instantons with radial symmetry. Two special
points where the Nahm zero modes have softer singularities are identified as
the locations of instanton core constituents. For a square torus
this constituent picture is closely reflected in the action density.
In rectangular tori with large aspect ratios the cores merge 
to form monopole-like objects. For particular values
of the parameters the torus can be cut in half yielding
two copies of a twisted charge ${1\over2}$ instanton. 
These findings are illustrated with plots of the action density
within a two-dimensional slice containing the constituents.

\end{abstract}
\baselineskip=18.9pt
\section{Introduction}

Monopoles are at the heart of various kinds of electric-magnetic
duality in extended super Yang-Mills theory, which have led to an
impressive body of quantitative results.  The role of monopoles and
other topological constructs in non-supersymmetric Yang-Mills theory
is less clear.  Using such objects, particularly monopoles and
vortices, models of confinement and chiral symmetry breaking have been
advanced.  A weakness of these models is that classical Yang-Mills
does not possess monopole solutions, rather they have instanton
solutions.  However, in recent years it has become clear that \it
periodic instantons \rm have a substructure that can be identified
with monopoles and other extended objects.  In particular, charge one
$SU(N)$ calorons (finite temperature instantons)
can be viewed as bound states of $N$ monopole
constituents \cite{Lee:1998bb,Kraan:1998pm}.  Provided the `size' of
the instanton is larger than the period (or inverse temperature) tubes
of action density form. These can be identified with the worldlines of
the constituent
monopoles.  It would be interesting to extend these results to
higher charge sectors and higher tori, i.e.  more than one period so
that we are considering instantons on $\T^n\times\R^{4-n}$, $n>1$.
Indeed, lattice based numerical studies indicate that doubly periodic
instantons ($n=2$) have remarkable properties including fractionally
charged solutions, vortex-like objects and an exponentially decaying
action density \cite{Gonzalez-Arroyo:1998ez}.  None of these
attributes are shared by the calorons. Nevertheless, it would be
desirable to have a unified description of calorons and
multiply-periodic instantons and their constituents.

The calorons are best understood within a formalism due to Nahm
\cite{Nahm:1979yw,Nahm:1983sv}.  This is an extension of the
 ADHM construction.  Nahm's approach was generalised to
other four manifolds, notably the four-torus, $\T^4$
\cite{Braam:1988qk}.  Here it becomes a duality, mapping $U(N)$ charge
$k$ instantons on the four torus to $U(k)$ charge $N$ instantons on
the dual torus, $\tilde\T^4$ (defined by inverting the four periods).
The caloron construction and even ADHM can be understood as a limiting
case of the four torus duality.  One therefore expects that doubly
periodic instantons can be treated in a similar fashion.
Jardim \cite{jardim} has used the Nahm transform to discuss the
charge one case.
  In
\cite{Ford:2000zt} doubly periodic charge one instantons with radial
symmetry in the non-compact $\R^2$ directions were considered. Under
the Nahm transformation they are mapped to abelian potentials on the
dual torus $\tilde\T^2$.  Starting with these rather simple Nahm
potentials the original instantons can be `recovered' via the
inverse Nahm transform.  Technically, this involves solving certain
Weyl-Dirac equations; for each $x\in\T^2\times\R^2$ one has a
different Weyl equation on $\tilde\T^2$. The Weyl zero modes were
determined explicitly for a two dimensional subspace (corresponding 
to the origin of $\R^2$).

In a recent letter \cite{Ford:2002pa} we used these zero modes to
compute the action density, analytically and numerically, within the
subspace.  We found that there are points in $\T^2\times\R^2$ where
the solution of the Weyl equations have softer singularities.  These
are obvious candidates for constituent locations.  We interpreted
these constituents as overlapping \it instanton cores.\rm\, For square
tori our constituent picture correlates closely to the action density.
Some of the charge one solutions we
considered can be cut in half yielding two copies of a charge
$\frac{1}{2}$ instanton.  This procedure is only consistent if the
gauge potential has specific asymptotic boundary conditions which are
analogous to a four torus twist.

In this paper we obtain further information about
doubly periodic instantons.
In particular, action density plots for rectangular tori
 are presented. These clearly show how doubly periodic instantons interpolate
between fractionally charged lumps characteristic of four torus solutions
and monopole like tubes of actions density typical of calorons.
Furthermore, we present an analytical discussion of the exponential
decay of the action density.  
The paper is organised as follows.  In section 2 the basics of the
Nahm transform are briefly reviewed, and its application to doubly
periodic instantons is discussed.  Sections 3 to 5 deal with the Nahm
potentials, zero modes and constituents corresponding to charge one
instantons with gauge group $SU(2)$. Section 6 describes some
technical results regarding Green's functions that crop up in field
strength calculations.  Our numerical action density computations are
detailed in section 7.  Section 8 considers charge $\frac{1}{2}$
solutions and the asymptotics of the gauge field and field strength
are discussed in section 9.

\section {Nahm Transform for Doubly Periodic Instantons}

Before we specialise to $\T^2\times\R^2$ let us briefly recall how the
Nahm transform is formulated on the four torus \cite{Braam:1988qk}
(for a string-theoretic approach see \cite{hori}).
Start with a self-dual 
anti-hermitian $SU(N)$ potential, $A_\mu(x)$,
on a euclidean four-torus with topological charge $k$.  A gauge field
on $\T^4$ is understood to be an $\R^4$ potential which is periodic
(modulo gauge transformations) with respect to $x_\mu\rightarrow
x_\mu+L_\mu \quad (\mu=0,1,2,3)$, the $L_\mu$ being the four periods
of the torus. Self-duality is the requirement that $F_{\mu\nu}=\frac{1}{2}
\epsilon_{\mu\nu\alpha\beta}F_{\alpha\beta}$,
and we have taken the convention $\epsilon_{0123}=-1$.
  The next step is to turn the $SU(N)$ instanton into a
$U(N)$ instanton by adding a constant $U(1)$ potential,
$A_\mu(x)\rightarrow A_\mu(x)-iz_\mu$, where the $z_\mu$ are real
numbers.  We can regard the $z_\mu$ as coordinates of the \it dual
torus \rm, $\tilde\T^4$, since the shifts $z_\mu\rightarrow
z_\mu+2\pi/L_\mu$ can be effected via \it periodic \rm $U(1)$ gauge
transformations.

Now consider the $U(N)$ Weyl operator
\begin{equation}
D_z(A)=\sigma_\mu D^\mu_z(A),~~~~
D_z^\mu(A)=\partial^\mu+A^\mu(x)-iz^\mu 
\end{equation}
with $\sigma_\mu=(1,i\tau_1,i\tau_2,i\tau_3)$ where the $\tau_i$ are Pauli
matrices.  Provided certain mathematical technicalities are met
$D^\dagger_z(A)=-\sigma_\mu^\dagger D_z^\mu(A)$ has $k$
square-integrable zero modes $\psi^i(x;z)$ with $i=1,2,...,k$. The
Nahm potential is defined as
\begin{equation}\label{nahm}
\hat A_\mu^{ij}(z)=\int_{T^4}d^4x\,
{\psi^i}^\dagger(x;z)\frac{\partial}{\partial z^\mu}
\psi^j(x;z).\label{basicnahm}
\end{equation}
Here the zero modes are taken to be orthonormal.
Remarkably,
$\hat A(z)$ is a $U(k)$ instanton on the dual torus with topological charge
$N$. 
The associated field strength can be written as follows
\begin{equation}\label{associateF}
\hat F^{ij}_{\mu\nu}(z)
=\int_{T^4}d^4x\int_{T^4}d^4x'\,
\psi^i{}^\dagger(x;z)\sigma_\mu (D_z^\dagger D_z)^{-1}
(x,x')\sigma_\nu^\dagger
\psi^j(x;z)-[\mu\leftrightarrow\nu].
\end{equation}
The self duality of $F_{\mu\nu}$
implies $D^\dagger D=-\sigma_0 D_\mu D_\mu$.
Accordingly, $(D^\dagger_z D_z)^{-1}(x,x')$ commutes with all
$\sigma_\mu$, from which the self duality of $\hat F$ follows.
The `original' gauge
potential, $A_\mu(x)$, can be recovered by Nahm transforming
$\hat A_\mu(z)$
\begin{equation}\label{t4inverse}
A_\mu^{pq}(x)=\int_{\tilde T^4} d^4 z
~\psi^p{}^\dagger(x;z)
\frac{\partial}{\partial x^\mu}
\psi^q(z;x),\end{equation}
where the
$\psi^p(z;x)$, $p=1,2,...,N$ are an orthonormal set of zero
modes of
$D_x^\dagger(\hat A)=-\sigma_\mu^\dagger D_x^\mu(\hat A)$
with
$D_x^\mu={\partial}/{\partial z_\mu}+\hat A^\mu(z)-ix^\mu$.
In other words the four torus Nahm transform is involutive.

Formally, one can obtain the $\T^2\times\R^2$ Nahm transform by taking
 two of the periods, say
$L_0$ and $L_3$, to be infinite.
Given an $SU(N)$ instanton periodic with respect to
$x_1\rightarrow x_1+L_1$, $x_2\rightarrow x_2+L_2$ its Nahm
transform is
\begin{eqnarray}
\hat A_\mu^{ij}(z)&=&\int_{T^2\times R^2} d^4x\, 
\psi^i{}^\dagger(x;z)\frac{\partial}{\partial z^\mu}\psi^j(x;z),
~~~~~~~\mu=1,2, \label{t2nahm}\nonumber
\\
\hat A_\mu^{ij}(z)&=&\int_{T^2\times R^2} d^4x\,
\psi^i{}^\dagger(x;z)ix_\mu \psi^j(x;z),
~~~~~~~\mu=0,3, 
\end{eqnarray}
where the $\psi^i(x;z)$ with $i=1,...,k$
are orthonormal zero modes of 
$D_z^\dagger(A)$.
Note that we may gauge $z_0$ and $z_3$ to zero.
By assumption the zero modes
$\psi^i(x;z)$ are square integrable.
This does not guarantee
that the integrals in (\ref{t2nahm}) exist.
However, the integrals only diverge at special
$z_\mu$ values
where the zero modes do not decay exponentially.
The number of these points turns out to be $N$,
i.e. it is determined by the gauge group of the instanton.
To summarise, the Nahm transform maps doubly periodic
$SU(N)$ instantons with topological charge $k$
to self dual $U(k)$ potentials on the dual
torus, $\tilde \T^2$,
with $N$ singularities.

The dimensionally reduced self-duality equations (or Hitchin equations)
take a particularly simple form in complex coordinates 
\begin{eqnarray}\label{complex}
y=z_1+iz_2, & & \bar y=z_1-i z_2,
\end{eqnarray} 
with derivatives $\partial_y=\h(\partial_{z_1}-i\partial_{z_2})$,
$\partial_{\bar y}=\h(\partial_{z_1}+i\partial_{z_2})$.
Combine $\hat A_1(z)$ and $\hat A_2(z)$ into a complex gauge potential,
$\hat A_y=\frac{1}{2}(\hat A_1(z)-i\hat A_2(z))$
and
$\hat A_{\bar y}=\frac{1}{2}(\hat A_1(z)+i\hat A_2(z))$,
and form a `Higgs' field out of the remaining
components,
$\Phi(z)=\h(\hat A_0(z)-i\hat A_3(z))$.
The self-duality equations read
\begin{equation}
[D_y,\Phi]=0,~~~~~~~
\hat F_{y\bar y}=[D_y,D_{\bar y}]=[\Phi,\Phi^\dagger].
\end{equation}
The reader should be aware that we will occasionally use both the
cartesian $z$ and complex $y$ coordinates in the same equation. The
Nahm transform (\ref{t2nahm}) is a straightforward limit of the $\T^4$
version.  The inverse transform is a different matter.  That is, given
a (singular) solution of the Hitchin equations on the dual torus,
$\tilde \T^2$, how does one recover the corresponding doubly periodic
instanton?  To our knowledge there is not a mathematically rigorous
treatment of this issue in the literature.  Our approach will be to
study specific simple solutions of the Hitchin equations and
investigate whether the number of normalisable zero modes of
$D_x^\dagger(\hat A)$ matches the number of singularities of $\hat
A(z)$.  When this is the case we can then ask whether the natural
analogue of (\ref{t4inverse}), namely
\begin{equation}\label{t2inverse}
A^{pq}_\mu(x)=\int_{\tilde T^2} d^2z\,
\psi^p{}^\dagger(z;x)\frac{\partial}{\partial x^\mu}
\psi^q(z;x),
\end{equation}
provides
 doubly periodic instantons.

\section{Charge One}

In the one-instanton case
the Nahm potential is abelian and the
Hitchin equations reduce to 
\begin{equation}
\hat F_{y\bar y}=0,~~~~\partial_y\Phi=0,\label{abhitchin}
\end{equation}
so that
$\Phi$ is anti-holomorphic in $y$.  As we shall see, $A_y$ can be
taken as holomorphic.  Since $A_y$ is periodic it is an elliptic
function.  It is well known that they have at least two singularities.
We expect the number of singularities to correspond to $N$.  Let us
stick to the simplest case $N=2$. Thus $\hat A_y$
will have two simple poles \footnote{The
Weyl equations for Nahm
  potentials with higher poles do not admit normalisable zero modes.}.
Consider the ansatz
\begin{equation}\label{ansatz}
\hat A_y=\partial_y\phi,~~~~~~~~
\hat A_{\bar y}=-\partial_{\bar y}\phi,
\end{equation}
which gives $\hat F_{y\bar y}=-2\partial_y\partial_{\bar y}\phi$.
It follows from \eq{abhitchin} that $\phi$ must be harmonic except
at two singularities.  A suitable $\phi$ satisfies
\begin{equation}
\left(
\partial_{z_1}^2+\partial_{z_2}^2\right)\phi(z)
=-2\pi\kappa\left[
\delta^2(z-\omega)-\delta^2(z+\omega)\right],
\end{equation}
where $\kappa$ is a constant and $\pm\omega$ are the positions of the
two singularities (we have used translational invariance to shift the
`centre of gravity' of the singularities to the origin).  The delta
functions should be read as periodic (with respect to $z_1\rightarrow
z_1 +2\pi/L_1$ and $z_2\rightarrow z_2+2\pi/L_2$).  Physically, the
Nahm potential describes two Aharonov-Bohm fluxes of strength $\kappa$
and $-\kappa$ threading the dual torus.  They must have equal and
opposite strength to ensure a periodic $\hat A$.  We may assume that
$\kappa$ lies between $0$ and $1$ since it is possible via a
(singular) gauge transformation to shift $\kappa$ by an integer amount
(under such a transformation the \sl total \rm flux through
$\tilde\T^2$ remains zero). One can write $\phi$ explicitly in terms
of Jacobi theta functions (see for example\ \cite{mumford})
\begin{equation}\label{phi}
\phi(z)=\frac{\kappa}{2}\left(
\log
\frac{\left|\theta\left(
(y+\omega_1+i\omega_2)\frac{L_1}{2\pi}
+\frac{1}{2}+\frac{iL_1}{2L_2},\frac{iL_1}{L_2}\right)\right|^2}
{\left|\theta\left(
(y-\omega_1-i\omega_2)\frac{L_1}{2\pi}
+\frac{1}{2}+\frac{iL_1}{2L_2},\frac{iL_1}{L_2}\right)\right|^2}
+\frac{iL_1 L_2 \omega_2}{\pi} (y-\bar y)-
2\omega_2 L_1
\right),
\end{equation}
where $y$ and $\bar y$ are the complex coordinates introduced in \eq{complex}. 
The theta function is defined as
\begin{equation}
\theta(w,\tau)=
\sum_{n=-\infty}^\infty
e^{i\pi n^2\tau+2\pi in w},
~~~~~~~\hbox{Im}\,\tau>0,\end{equation}
and has the periodicity properties
$
\theta(w+1,\tau)=\theta(w,\tau)$ and
$\theta(w+\tau,\tau)=e^{-i\pi \tau-2\pi i w}
\theta(w,\tau)$.
In each cell $\theta(w,\tau)$ has a single zero
located at the centre of the torus ($w=\h+\h\tau$).
We have chosen the constant term in (\ref{phi}) so that
the integral of $\phi$ over the dual torus is zero.
This renders $\phi(z)$ an odd function, $\phi(-z)=-\phi(z)$.
It is easy to see that inserting (\ref{phi}) into
(\ref{ansatz}) yields a holomorphic $A_y$ with simple poles
at the fluxes.

What about the Higgs field?
Since we are aiming for an $SU(2)$ instanton it must, like $\hat A(z)$,
have two singularities. The singularities must be at the same
two positions as in $A_{\bar y}$ or else we have a total of four singularities
which means we are considering a (rather special) $SU(4)$ instanton.
The simplest way to arrange for $A_{\bar y}$ and the
Higgs to have 
singularities at the same points is to choose them
to be proportional 
\begin{equation}
\kappa\Phi=\alpha \partial_{\bar y} \phi,
\end{equation}
where $\alpha$ is a complex constant.  Not all possibilities are
exhausted, since while the poles must coincide the zeroes need not.
This remaining ambiguity corresponds to the freedom to add to $\Phi$ a
complex constant.  However when we insert our Nahm potential into the
Weyl equation such a shift is equivalent to a translation of $x_0$ and
$x_3$.  This Nahm potential was derived in \cite{Ford:2000zt} via the
ADHM construction.  Here the parameters $\kappa$ and $\alpha$ are
related to the `size', $\lambda$ of an instanton centred at $x_\mu=0$,
\begin{equation}\label{adhm}
\sqrt{\kappa^2+|\alpha|^2}=\frac{\pi \lambda^2}{L_1 L_2}.
\end{equation}
Although we do not directly use the ADHM construction in the present
paper this relation proves useful in interpreting our results.  The
Nahm potential we have described has five parameters; $\omega _1$,
$\omega_2$ fixing the flux separations, $\kappa$ the flux strength and
the complex parameter $\alpha$ specifying the Higgs field. Then, with
the four translations in $\T^2\times \R^2$, the corresponding doubly
periodic instanton has a total of nine parameters. 

To reconstruct the charge one $SU(2)$ instantons corresponding
to our simple Nahm potential we require the two zero modes of
 the Weyl operator
\begin{equation}\label{weylop}
-\frac{i}{2} D_x^\dagger(\hat A)=
\left(
\begin{array}{cc}
\h \bar x_\perp +i\alpha\kappa^{-1}\partial_{\bar y}\phi&
\partial_y+\partial_y\phi-\frac{i}{2}\bar x_{||}
\\
\partial_{\bar y}-\partial_{\bar y}\phi-\frac{i}{2}x_{||}
&\h x_\perp-i \bar\alpha\kappa^{-1}\partial_y\phi
\end{array}\right). 
\end{equation}
In addition to the complex coordinates $y$, $\bar y$ on $\tilde\T^2$
we have introduced two sets of complex coordinates
for $\T^2\times\R^2$; in the `parallel' directions
$x_{||}=x_1+ix_2$, $\bar x_{||}=x_1-ix_2$,
and in the `transverse' non-compact directions
$x_\perp=x_0+ix_3$, $\bar x_\perp=x_0-ix_3$.
In this paper we shall concentrate on
the special case $\alpha=0$, i.e. zero Higgs.
This means that the corresponding instanton will be radially symmetric:
the action density depends on $x_1$, $x_2$
and $r=|x_\perp|$ only. These radially symmetric solutions
are a seven parameter subset of the doubly periodic $SU(2)$
one-instantons.

\section{Nahm Zero Modes}

From now on we stick to the zero Higgs case (i.e.\ $\alpha=0$).
Even with this specialisation the Weyl equations
are still
rather forbidding.
However, there is one special case which
is immediately tractable:
when  $x_\perp=0$ the Weyl equation
decouples and the two zero modes have a simple form \cite{Ford:2000zt}
\begin{equation}\label{modes}
\psi^1(z;x)=\left(\begin{array}{cr}
0 \\ e^{-\phi(z)}G_+(z-\omega)
\end{array}\right),~~~~
\psi^2(z;x)=\left(\begin{array}{cr}
e^{\phi(z)}G_-(z+\omega)\\
0
\end{array}\right),
\end{equation}
where $G_\pm(z)$ are periodic Green's functions satisfying
\begin{equation}
\left(-i\partial_{y}-\h\bar x_{||}\right)G_+(z)
=\h\delta^2(z),~~~~
\left(-i\partial_{\bar y}-\h x_{||}\right)
G_-(z)=\h\delta^2(z).
\end{equation}
Here we have used both the $z$ 
and $y$ coordinates in the
 same equation; this possibly confusing usage will continue until the end of 
section~\ref{sec:soft}. 
 $G_-(z)$ has a theta function representation
\begin{equation}\label{szeboe}
G_-(z)=\frac{iL_1}{4\pi^2}
e^{\h ix_{||}(\bar y-y)}
\frac{\theta'(\h+\frac{iL_1}{2L_2},\frac{iL_1}{L_2})\,\theta
(\frac{L_1}{2\pi} 
y +\h+\frac{iL_1}{2L_2}-\frac{i}{L_2}x_{||},\frac{iL_1}{L_2})}{
\theta(\h+\frac{iL_1}{2L_2}-\frac{i}{L_2}x_{||},\frac{iL_1}{L_2})\,
\theta(\frac{L_1}{2\pi} y+\h+\frac{iL_1}{2L_2},\frac{i L_1}{L_2})},
\end{equation}
with $\theta'(w,\tau)=\partial_w \theta(w,\tau)$.
The corresponding result for
$G_+(z)$ can be obtained via  $G_+(z)=G_-^*(-z)$.
The zero modes have square-integrable singularities at both fluxes:
near $z\sim\omega$ we have 
$|\psi^1|^2\propto|y-\omega_1-i\omega_2|^{2(\kappa-1)}$
and
$|\psi^2|^2\propto
|y-\omega_1-i\omega_2|^{-2\kappa}$.
Near the other flux the singularity profiles of the two modes are exchanged
 in that
$|\psi^1|^2\propto|y+\omega_1+i\omega_2|^{-2\kappa}$
and
$|\psi^2|^2\propto
|y+\omega_1+i\omega_2|^{2(\kappa-1)}$.
By virtue of the second theta function in the numerator of
(\ref{szeboe}) $\psi^2(z;x)$ has a single zero in $\tilde \T^2$
at $y=-\omega_1-i\omega_2-2\pi i x_{||}/(L_1L_2)
$ (see also \cite{Reinhardt:2002cm}).
Similarly, $\psi^1(z;x)$ has a zero at $y=
\omega_1+i\omega_2-2\pi i x_{||}/(L_1L_2)$.

\section{Soft Zero Modes}\label{sec:soft}

When $x_{||}=0$ the Green's functions $G_\pm$ do not exist.  This does
not mean there are no zero modes, indeed one can see that
\begin{equation}
\psi^1(z;x=0)=
\left(\begin{array}{cr}
0\\e^{-\phi(z)}\end{array}\right),~~~~~~~
\psi^2(z;x=0)=
\left(\begin{array}{cr}
e^{\phi(z)}\\
0
\end{array}\right),
\end{equation}
are solutions of the $x_{||}=0$ Weyl equation.  $\psi^1$ has the
expected square-integrable singularity at $z=-\omega$, but for
$z=\omega$, $\psi^1(z;x=0)$ is zero.  On the other hand
$\psi^2(z,x=0)$ diverges at $z=\omega$ but not at $z=-\omega$. By
choosing $x_{||}=0$ we have effectively moved the zeroes of the zero
modes to flux locations.  There is another way of doing
this; choosing $x_{||}=-iL_1 L_2(\omega_1+i\omega_2)/\pi$ brings the zero
of $\psi^1$ to the other flux (and similarly for $\psi^2$).  Thus
there are two $x_{||}$ values where each Nahm zero mode diverges at
only one flux. We wish to investigate whether these `soft' points can
be interpreted as the locations of some kind of constituent.

The singularity profiles of $\psi^1$ and $\psi^2$ are exchanged under
the replacement $\kappa\rightarrow 1-\kappa$ suggesting that the
constituents are exchanged under this mapping. That is, if there are
indeed lumps at the two points, then $\kappa\rightarrow 1-\kappa$
swaps the two lumps.  The following result formalises this idea
\begin{equation} \label{duality}
F_{\mu\nu}(x_{||},x_\perp,\kappa)=
V^{-1}(x)
F_{\mu\nu}\left(-x_{||}+\frac{L_1 L_2}{i\pi}(\omega_1+i\omega_2),-x_\perp,
1-\kappa\right)V(x),
\end{equation}
where $V(x
)$ is some $U(2)$ gauge transformation. 
The proof of \Eq{duality} goes as follows: make the change of variables
$z\rightarrow -z$ in (\ref{t2inverse}). The zero mode $\psi^p(-z;x)$
satisfies the same Weyl equation as $\psi^p(z;x)$ except
that the signs of $\kappa$ and the $x_\mu$ are flipped.
However, $-\kappa$ does not lie between $0$ and $1$.
Under periodic gauge
transformations 
$\hat A(z,-\kappa)$ and $\hat A(z,1-\kappa)$
 are equivalent up to
a constant potential 
\begin{equation}
\hat A_\mu(z,-\kappa)= U^{-1}(z)\left(
\partial_{\mu}+\hat A_\mu(z,1-\kappa)+B_\mu\right)U(z), 
\end{equation}
where $\pi B_1=-iL_1 L_2\omega_2$ and $\pi B_2=iL_1 L_2 \omega_1$.
Explicitly
\begin{eqnarray}
U(z)&=&
e^{\h\omega_1 L_1 L_2(y-\bar y)/\pi}\\ 
&\times&\left[
\frac{
\theta\left(\frac{L_1}{2\pi}(\bar y-\omega_1+i\omega_2)+\h-\frac{iL_1}{2L_2}
,\frac{iL_1}{L_2}\right)
\theta\left(\frac{L_1}{2\pi}( y+\omega_1+i\omega_2)+\h+\frac{iL_1}{2L_2}
,\frac{iL_1}{L_2}\right)}
{\theta\left(\frac{L_1}{2\pi}( y-\omega_1+i\omega_2)+\h+\frac{iL_1}{2L_2}
,\frac{iL_1}{L_2}\right)
\theta\left(\frac{L_1}{2\pi}( \bar y+\omega_1-i\omega_2)+\h-\frac{iL_1}{2L_2}
,\frac{iL_1}{L_2}\right)}\right]^{\h}.\nonumber
\end{eqnarray}
In the Weyl equation $B_\mu$ can be absorbed into $x_1$ and
$x_2$.
Thus $U^{-1}(z)\psi^p(z,-x_{||}-i L_1 L_2(\omega_1+i\omega_2)/\pi,
-x_\perp)$ satisfies the same Weyl equation as
$\psi^p(-z;x_{||},x_\perp)$.

\section{Green's Function Approach}\label{Greens}

 For the $\T^4$ transform, the original field
strength can be written
\begin{equation}\label{t4strength}
F_{\mu\nu}^{pq}(x)=\int_{\tilde T^4}d^4z
\int_{\tilde T^4} d^4z'\, 
\psi^p{}^\dagger(z;x)\sigma_\mu~(D_x^\dagger D_x)^{-1}(z,z')~\sigma_\nu^\dagger
\psi^q(z';x)-[\mu\leftrightarrow\nu],\end{equation}
and as with the corresponding expression for $\hat F_{\mu\nu}^{ij}
(z)$, equation \eq{associateF}, the self-duality of $F_{\mu\nu}^{pq}(x)$
is manifest because the Green's function $(D_x^\dagger D_x)^{-1}(z,z')$
commutes with all $\sigma_\mu$.
Starting directly from the inverse Nahm transform (\ref{t4inverse}) we have
\begin{eqnarray}\label{direct}
F_{\mu\nu}^{pq}(x)&=&
\int_{\tilde T^4}d^4z\,
\partial_\mu\psi^p{}^\dagger(z;x)
\partial_\nu\psi^q(z;x)\\ \nonumber
&&+\int_{\tilde T^4}d^4z\,
\psi^p{}^\dagger(z;x)\partial_\mu\psi^r(z;x)
\int_{\tilde T^4} d^4z'\,
\psi^r{}^\dagger(z';x)\partial_\nu\psi^q(z';x)
-[\mu\leftrightarrow\nu], 
\end{eqnarray}
where $\partial_\mu=\partial/\partial x^\mu$. To get from 
(\ref{direct}) to (\ref{t4strength})
we require the derivative of the Nahm zero modes with respect to the 
coordinates of $\T^4$
\begin{equation}\label{modederivative}
\frac{\partial}{\partial x^\mu}\psi^q=- i\,
D_x(\hat A)\left(
D_x^\dagger(\hat A)
D_x(\hat A)\right)^{-1}\sigma_\mu^\dagger\psi^q+
\psi^r~R_\mu^{rq}(x),\end{equation}
for some $N\times N$ matrix valued
vector field on the four torus, $R_\mu^{qr}(x)$.
This clearly satisfies the equation obtained
by differentiating the Weyl equation, $D_x^\dagger(\hat A)\psi^q(z;x)=0$,
with respect to $x^\mu$.
Multiplying \eq{modederivative} from the left with $\psi^p{}^\dagger$ and 
integrating over the dual torus yields
\begin{equation}\label{t4R}
R^{pq}(x)=A^{pq}(x),
\end{equation}
which together with equations (\ref{direct}) and 
(\ref{modederivative}) give (\ref{t4strength}).

Our analysis of doubly periodic instantons
has been based on the Weyl equation 
and the inverse Nahm transform (\ref{t2inverse}).
These are the exact counterparts of  four torus
equations.
Therefore one might expect all the results quoted above
to carry over directly to $\T^2\times\R^2$.
There are, however, a number of important differences between the two cases.
With regard to the $\T^2\times\R^2$
construction, 
the Green's function $(D_x^\dagger(\hat A)D_x(\hat A))^{-1}(z,z')$
does not commute with the $\sigma_\mu$ matrices.
Away from the $N$ singularities of
$\hat A(z)$,
$D_x^\dagger(\hat A)D_x(\hat A)$ commutes with the $\sigma_\mu$.
At the singularities it has source terms which are not
proportional to $\sigma_0$.
For the $U(1)$ Nahm potential considered in section
3 (as usual we set $\alpha=0$)
\begin{equation}D_x^\dagger D_x=
-\sigma_0 (D_x^\mu)^2+2\pi i \kappa\sigma_3\left[
\delta^2(z-\omega)-\delta^2(z+\omega)\right],
\end{equation}
indicating that $(D_x^\dagger D_x)^{-1}$ does not commute with
$\sigma_1$ and $\sigma_2$.

We will show that the naive
$\T^2\times\R^2$ limit of (\ref{t4strength}) does not apply, so that
$(D_x^\dagger D_x)^{-1}$ 
not being proportional to $\sigma_0$ does not signal
a breakdown of the inverse Nahm transform.  Starting with the inverse
transform (\ref{t2inverse}), the field strength can be written in
exactly the same way as in (\ref{direct}) except that the integral is
over $\tilde \T^2$ rather than $\tilde\T^4$.  The derivative formula
(\ref{modederivative}) also applies without modification to the
$\tilde\T^2$ zero modes.  What does not carry through is equation
(\ref{t4R}).  Much as in the four torus analysis $R_\mu$ can be determined
by multiplying the $\T^2\times \R^2$ analogue of \eq{modederivative} by
$\psi^p{}^\dagger$ and integrating over $\tilde \T^2$.
\begin{equation}\label{t2R}
A_\mu^{pq}(x)
=-i \int_{\tilde T^2}  
\psi^p{}^\dagger 
D_x(\hat A)\left(D_x^\dagger(\hat A)D_x(\hat A)\right)^{-1}
\sigma^\dagger_\mu\psi^q +R_\mu^{pq}.
\end{equation}
Naively one would expect the integral on the right hand side to vanish
as $\psi^p{}^\dagger$ is a (left) zero mode of $D_x(\hat A)$. However,
this property only leads to vanishing $\int_{\tilde T^2}
\psi^p{}^\dagger D_x(\hat A) \phi$ for sufficiently smooth functions $\phi$.
 In \eq{t2R} 
$\psi^p{}^\dagger$ and the Green's function $(D_x^\dagger D_x)^{-1}$
have coincident singularities. 

Retaining the integral yields a modified field strength formula
\begin{equation}\label{fF}
F_{\mu\nu}^{pq}(x)
=\int_{\tilde T^2}d^2z\int_{\tilde T^2}d^2z'\,
\psi^p{}^\dagger(z;x)\sigma_\mu\,
f(z,z';x)\,\sigma_\nu^\dagger\psi^q(z';x)-[\mu\leftrightarrow\nu],
\end{equation}
with
\begin{eqnarray}\label{fdef}
f(z,z';x)
&=&\left(D_x^\dagger D_x\right)^{-1}(z,z')
\\ \nonumber
&&-\int_{\tilde T^2}
d^2s\left(D_x^\dagger D_x\right)^{-1}(z,s)
\,D_x^\dagger(\hat A)\psi^p(s;x)\\ \nonumber
&&~~~~~~~~\times\int_{\tilde T^2}
d^2s'\,
\psi^p{}^\dagger(s';x)D_x(\hat A)\,
\left(D_x^\dagger D_x\right)^{-1}(s',z').
\end{eqnarray}
This object does commute with the $\sigma_\mu$ consistent with
self-duality.  The field strength formula \eq{fF} is
similar to those one can derive for $\R^4$ instantons and calorons via
the ADHM formalism.

To make these considerations more concrete we see how they apply to our
simple $U(1)$ Nahm potential. In this case
\begin{equation}
\frac{1}{4}D_x^\dagger D_x=
\left(\begin{array}{cr}
\frac{1}{4}|x_\perp|^2+
(-\partial_y-\partial_y\phi+\frac{i}{2}\bar x_{||})
(\partial_{\bar y}-\partial_{\bar y}\phi-\frac{i}{2}x_{||})~~~~~~~~~~~0~~~~~
~~~~~~~~\\
~~~~~~~~~~~~~0~~~~~~~~~~~
\frac{1}{4}|x_\perp|^2+(-\partial_{\bar y}+\partial_{\bar y}
\phi+\frac{i}{2}x_{||})
(\partial_y+\partial_y\phi-\frac{i}{2}\bar x_{||})
\end{array}\right).
\end{equation}
When $x\neq0$ this is invertible. The inverse has the form
\begin{equation}\label{ddagdinverse}
2\left(D_x^\dagger D_x\right)^{-1}(z,z')=
(\sigma_0+i\sigma_3)e^{-\phi(z)}K_+(z,z';x)e^{-\phi(z')}+
(\sigma_0-i\sigma_3)e^{\phi(z)}K_-(z,z';x)e^{\phi(z')},
\end{equation}
where the $K_\pm(z,z';x)$ are finite for all $z$ and $z'$
(unless $x_{||}=x_\perp=0$).
Because of the exponentials
 $(D_x^\dagger D_x)^{-1}(z,z')$ is singular if $z$ or $z'$
equals $\pm\omega$.
In the $x_\perp=0$ slice the $K_\pm$ have simple integral representations
\begin{equation}\label{kxperpzero}
K_\pm(z,z';x_\perp=0)=
\int_{\tilde T^2}d^2 s\,G_\pm(z-s)e^{\pm2\phi(s)}G_\mp(s-z'), 
\end{equation}
where  $G_\pm$ are the Green's functions defined in section 4. The 
$x_\perp=0$ zero modes of section 4 can be rewritten as follows 
\begin{eqnarray}\label{zeromodes}
\psi^1(z;x)& =& -D_x(\hat A)\left(\begin{array}{cr}
e^{\phi(z)}K_-(z,\omega;x)\\
0\end{array}\right)\\[1mm] \nonumber
\psi^2(z;x)&= & -D_x(\hat A)\left(\begin{array}{cr}
0\\
e^{-\phi(z)}K_+(z,-\omega;x)\end{array}\right).
\end{eqnarray}
In fact, these are also valid for $x_\perp\neq0$.  The modes are
orthogonal and can be normalised by dividing them by $\sqrt {\rho}$,
where
\begin{equation}
\rho(x)=K_+(-\omega,-\omega;x)=K_-(\omega,\omega;x).\end{equation}
Inserting the zero modes into (\ref{fdef}) enables us to prove that 
$f$ commutes with $\sigma_1, \sigma_2$. This is detailed in 
appendix~\ref{app:proof}. One can also show that $f(z,z';x)$, 
unlike $(D_x^\dagger D_x)^{-1}(z,z')$,
is well behaved at $x=0$. As $z$ or $z'$ approaches $\pm\omega$,
$f$ tends to zero.

We have seen that it is possible to construct the Nahm zero modes out
of objects, $K_\pm$, entering the inverse of $D_x^\dagger D_x$. The
gauge potential can be computed from these zero modes. The final
result is very simple
\begin{eqnarray}\label{fullpot}
A_{x_{||}}&=&-\frac{\tau_3}{2}
\partial_{x_{||}}\log\rho-2\pi i(\tau_1-i\tau_2)\kappa\rho
\partial_{\bar x_\perp}\frac{\nu^*}{\rho},\\ \nonumber
A_{x_\perp}&=&-\frac{\tau_3}{2}
\partial_{x_{\perp}}\log\rho+2\pi i(\tau_1-i\tau_2)\kappa\rho
\partial_{\bar x_{||}}\frac{\nu^*}{\rho},
\end{eqnarray}
where
\begin{equation}
\label{defofnu}
\nu(x)=K_-(\omega,-\omega;x),~~~~~~~~~~~~
\nu^*(x)=K_-(-\omega,\omega;x).
\end{equation}
Note that $\rho$ is dimensionless, real and periodic (with respect to
$x_1\rightarrow x_1+L_1$, $x_2\rightarrow x_2+L_2$), while $\nu$ is
dimensionless, complex and periodic up to constant phases.
\Eq{fullpot} can be checked component-wise (see 
appendix~\ref{app:outline}).

We have given  closed forms for the $K_\pm$ in the special
case  $x_\perp=0$. An explicit construct valid beyond this 
two-dimensional slice would immediately provide the exact
Nahm zero modes and gauge potential. Note that the Green's functions
$K_\pm$ 
are radially symmetric.
Therefore the functions $\rho$ and $\nu$ are as well.
They can be expressed as power series
\begin{equation}\label{series}
\rho=\sum_{n=0}^\infty\rho_n|x_\perp|^{2n},~~~~~~~~~
\nu=\sum_{n=0}^\infty\nu_n|x_\perp|^{2n}.\end{equation}

\section{ Field Strength Components}

In section 5 we
saw that the Nahm transform singled out two points in the $x_\perp=0$
slice. This suggests that they may be the locations of some kind of 
constituent. As
 they are
 isolated points
 in $\T^2\times \R^2$ rather than lines it appears that we are \it not \rm
dealing with monopole constituents.
Indeed if we take (\ref{adhm}) at face value 
 the constituents appear to be BPST
instantons,
one with size
$\lambda=\sqrt{\kappa L_1 L_2/\pi}$
at $x_{||}=x_\perp=0$,
the other with size
$\lambda=\sqrt{(1-\kappa) L_1 L_2/\pi}$
at the second soft point.
This description shares some features with early attempts to describe
$SU(2)$ instantons in terms of $2k$ `instanton quarks' \cite{Belavin:fb}.

To see if this is realised  we must compute the field strengths.
This can be done  explicitly in the $x_\perp=0$ slice.
Inserting the power series (\ref{series}) into the ansatz
(\ref{fullpot}) it is straightforward to compute
$F_{x_{||}\bar x_{||}}$ and $F_{x_{||}\bar x_\perp}$
\begin{equation}
\left.F_{x_{||}\bar x_{||}}\right|_{x_\perp=0}=
\tau_3\partial_{x_{||}}\partial_{\bar x_{||}}
\log\rho_0,~~~~~~~~~
\left.F_{x_{||}\bar x_\perp}\right|_{x_\perp=0}
=2\pi i(\tau_1+i\tau_2)\,
\kappa\rho_0\,\partial^2_{x_{||}}\frac{\nu_0}{\rho_0}. 
\end{equation}
The other components are fixed by 
 self-duality, i.e. $F_{x_{||}\bar x_{||}}+F_{x_\perp\bar x_\perp}=0$
and $F_{x_{||}x_\perp}=0$.
 Both  $\rho_0$ and $\nu_0$ diverge at
$x_{||}=0$, but the field strengths should not.
A careful analysis of $\rho_0$ and $\nu_0$ in the neighbourhood of
$x_{||}=0$ shows that the field strengths are well defined at this point.
Alternatively, one can just note that $\rho$ and $\nu$ are well behaved
at the second soft point
and invoke (\ref{duality}).
Using these results
 the action density can be computed
\begin{equation}
\label{actiondensity}
-\h \hbox{Tr} \left.F_{\mu\nu}F^{\mu\nu}\right|_{x_\perp=0}=
16\left(\partial_{x_{||}}\partial_{\bar x_{||}}\log\rho_0
\right)^2+
128\pi^2\kappa^2\rho^2_0\left|
\partial_{x_{||}}^2\frac{\nu_0}{\rho_0}\right|^2.
\end{equation}
From (\ref{kxperpzero}) one can derive
integral representations of $\rho_0$ and $\nu_0$ where
the integrands are expressible in terms of standard functions.
These representations allow us to plot the action density within the
$x_\perp=0$ slice for different values of the parameters
$\kappa$, $\omega_1$ and $\omega_2$. The two
periods $L_1$ and $L_2$ can also be varied. However,
 the physically important 
quantity is the ratio of the lengths
\begin{equation}
a=\frac{L_1}{L_2},
\end{equation}
since scaling identically both lengths equates to a trivial scaling
of the action density:
\begin{eqnarray}\label{ident} 
\beta^4\,
\Tr\, F^2(\beta x;\,\beta L_1, \beta  L_2 ;\,\beta^{-1} {\omega_1}, 
\beta^{-1}{\omega_2};\, \kappa)= 
\Tr\, F^2(x;\,L_1,L_2;\,\omega_1,\omega_2;\, \kappa). 
\end{eqnarray}
In \cite{Ford:2002pa} plots for different $\kappa$ in the equal length
case, $a=1$, were presented.  There we found that the action density
has either one or two peaks.  A single peak is observed when the two
soft points are quite close or when one size is somewhat larger than the
other. In the latter case the smaller sized instanton dominates. Even
when two peaks are resolved there is quite a strong overlap. In fact
such an overlap is `necessary' since otherwise the instanton would
have charge two rather than one.  The constituent locations then
correspond to the `cores' of the overlapping instantons.

Here we also consider rectangular tori with $a\neq 1$. Without loss
of generality we set $L_2=1$ and $L_1=a$.  To begin with consider
$\kappa=\frac{1}{2}$ where the two constituents are identical. Now
choose $(\omega_1,\omega_2)=(\frac{1}{2}\pi a^{-1},\frac{1}{2}\pi)$.
This means that there is a core at the centre of the torus and another
of equal size at the corners.  Starting with $a=1$ increase the value
of $a$ thereby stretching the torus in the $x_1$ direction.

\begin{figure}[h]
\begin{picture}(0,400)(-250,0)
\put(-250,230){\epsfig{file=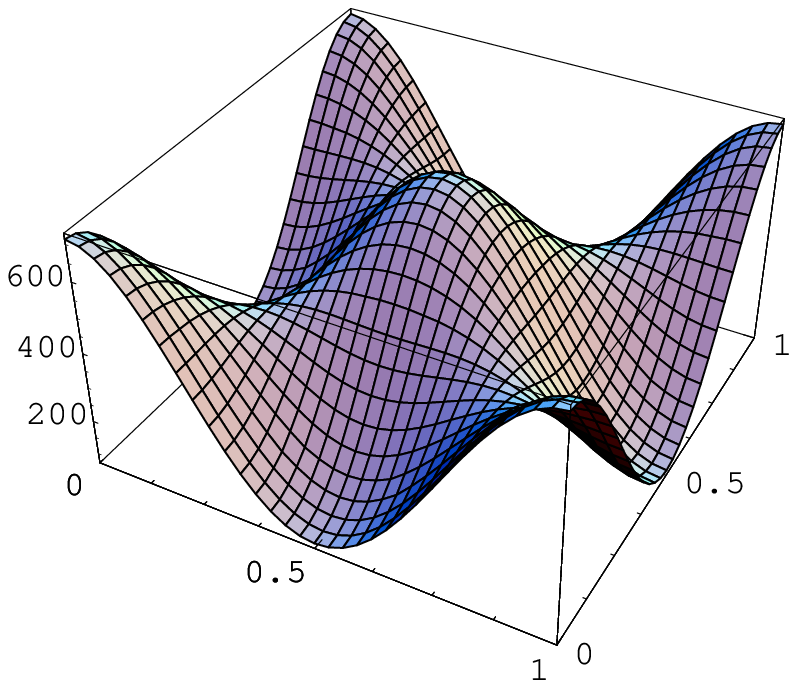,width=.4\hsize}}
\put(-250,375){$a=1$} 
\put(-202,232){$\bf  x_2$}  
\put(-65,266){ $\bf x_1$}
\put(0,230){\epsfig{file=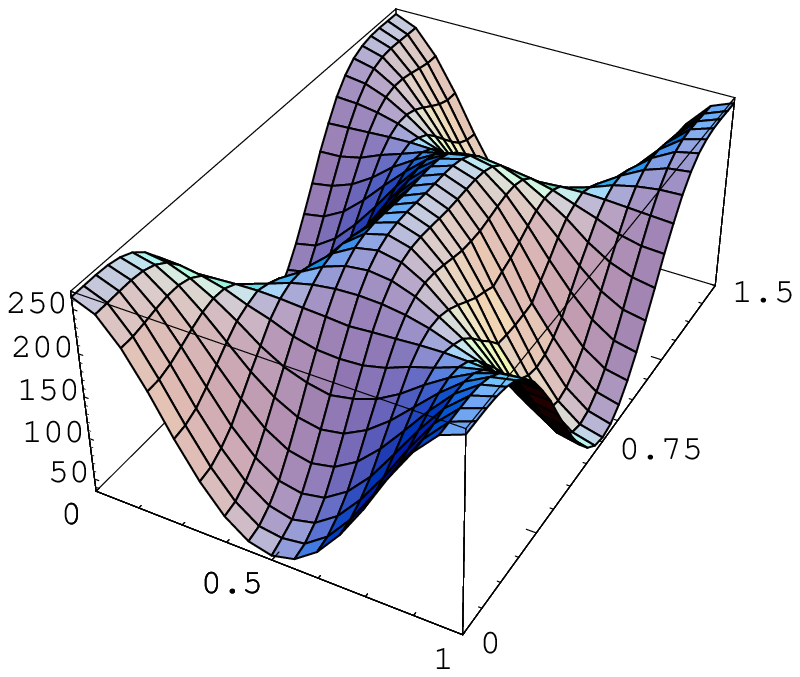,width=.4\hsize}}
\put(10,375){$a=\s032$} 
\put(48,228){$\bf  x_2$}  
\put(175,269){ $\bf x_1$}
\put(-250,20){\epsfig{file=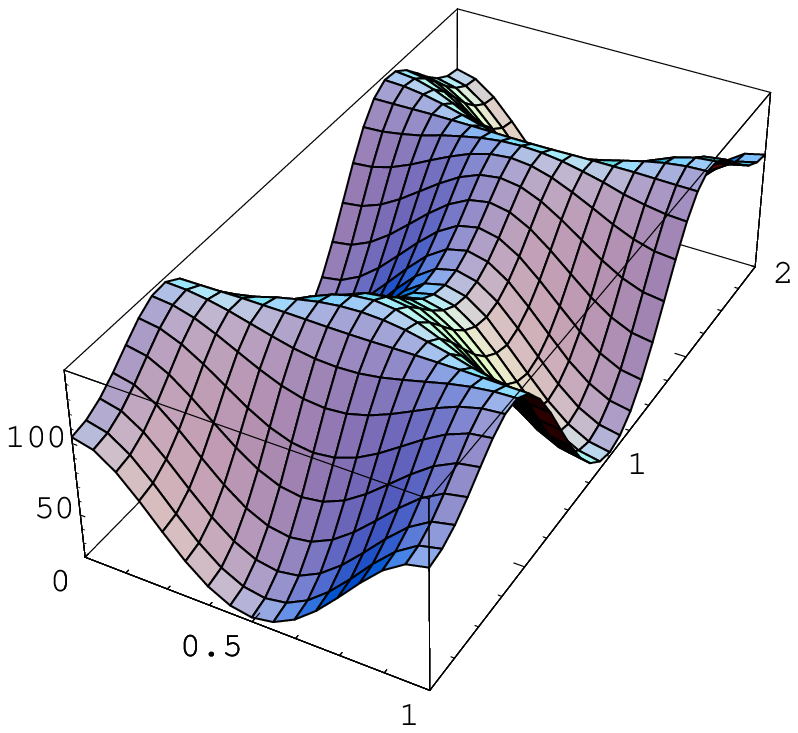,width=.4\hsize}}
\put(-240,180){$a=2$} 
\put(-207,14){$\bf  x_2$}  
\put(-75,76){ $\bf x_1$}
\put(0,20){\epsfig{file=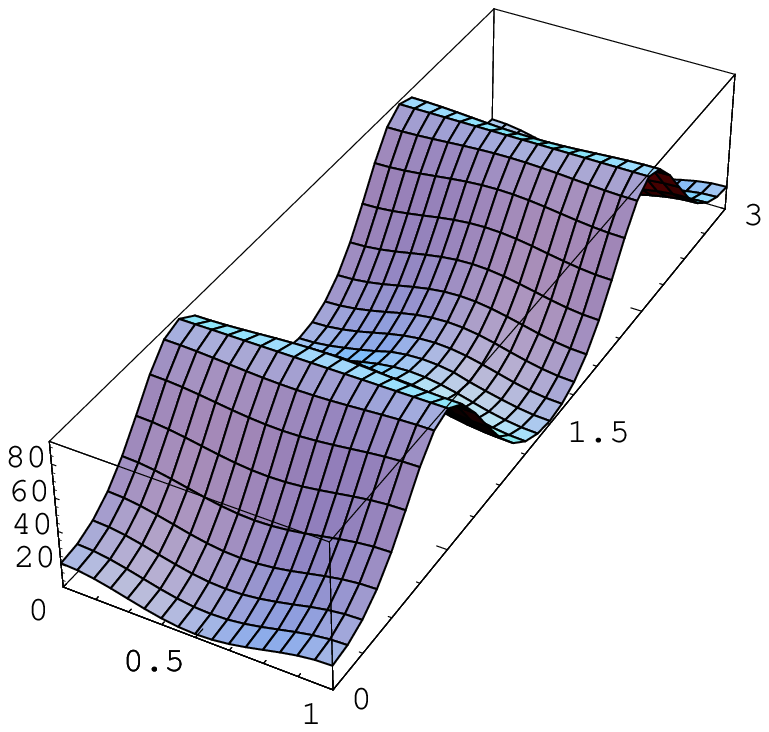,width=.4\hsize}}
\put(10,180){$a=3$} 
\put(31,13){$\bf  x_2$}  
\put(175,76){ $\bf x_1$}
\end{picture}
\caption{action density $-\s012 \Tr\, F^2$ for $x_\perp=0$,  
$\kappa=\s012$  and $\omega_1=\s012\pi, \omega_2=\s012 \pi a$ for 
$a=1,\s032,2,3$.}
\label{history}
\end{figure}

In fig.~\ref{history} the action density is plotted for some values of
$a$ ranging from $1$ to $3$.  The first plot ($a=1$) is just the equal
length case clearly showing two instanton-like peaks exactly at the
expected locations.  The second plot ($a=\frac{3}{2}$) shows that as
the torus is stretched the two cores become stretched as well; they no
longer have the approximate spherical symmetry of the $a=1$ case.  On
stretching the torus further the now deformed cores seek to overlap
with the periodic copies of themselves.  This has already started to
happen in the third plot where $a=2$; the soft point
locations are no longer maxima of the action density.  Finally, at
$a=3$ the self-overlap has generated two monopole like objects; the
solution has a very weak dependence on $x_2$ and there are peaks at
two values of $x_1$.  Note that the two monopole `worldlines' do not
cross the soft points.  Increasing $a$ more weakens further the
dependence on $x_2$. In the limit $a\rightarrow\infty$ a periodic
monopole (i.e. a self dual solution on $S^1\times\R^2$) will form.
However, this is a singular limit.

\begin{figure}[h]
\begin{picture}(0,250)(-250,20)
\put(-250,30){\epsfig{file=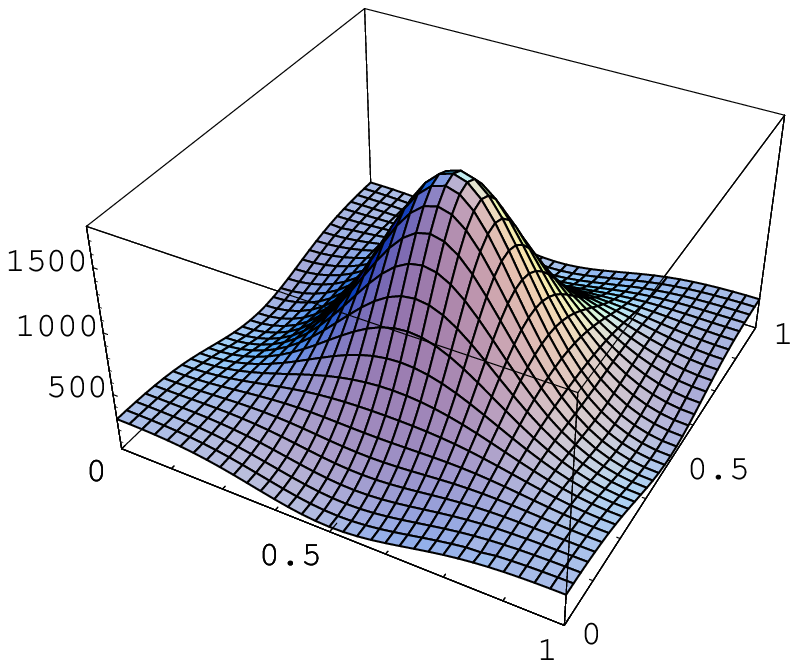,width=.4\hsize}}
\put(-245,170){$a=1$} 
\put(-190,28){$\bf  x_2$}  
\put(-63,60){ $\bf x_1$}
\put(0,30){\epsfig{file=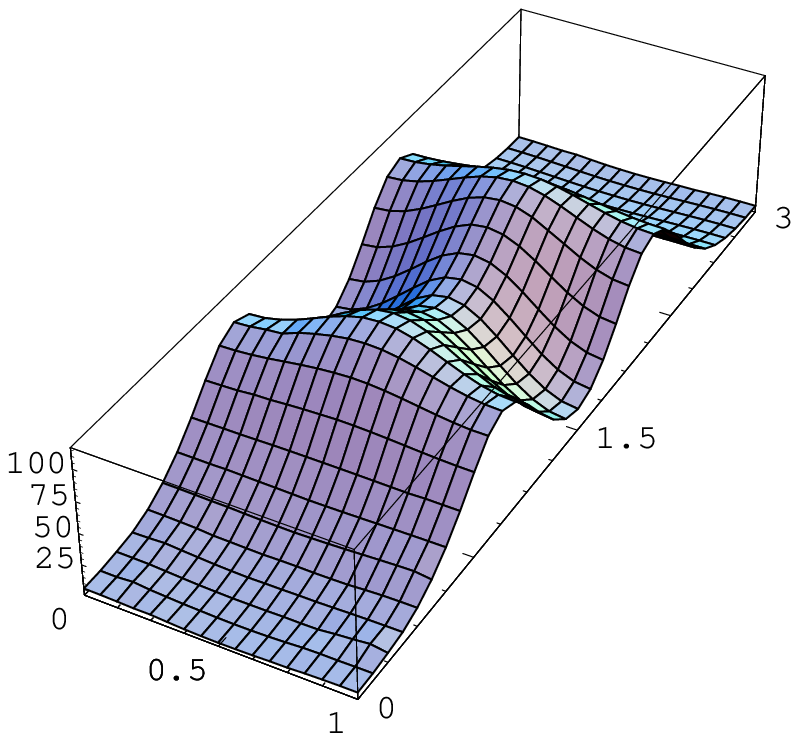,width=.4\hsize}}
\put(5,170){$a=3$} 
\put(39,30){$\bf  x_2$}  
\put(170,80){ $\bf x_1$}
\end{picture}
\caption{action density $-\s012 \Tr\, F^2$ for $x_\perp=0$,  
$\kappa=\s058$  and $\omega_1=\s012\pi, \omega_2=\s0{1}{2}\pi a$ with $a=1,3$.}
\label{3over8}
\end{figure}

What is happening is that the doubly periodic instantons are becoming
more caloron-like (caloron-like objects have been observed recently in
lattice-based approaches \cite{Gattringer:2002tg,Ilgenfritz:2002qs} ).
It is helpful to recall some of the basic features of the simplest
caloron solutions. An SU(2) caloron with charge one comprises two
monopole constituents separated by a distance $\pi\lambda^2/L_2$
(taking $x_2$ to be the compact coordinate).  Here $\omega_2$ (there
is no $\omega_1$) determines the mass ratio of the two monopoles.  For
the $\T^2\times\R^2$ problem $\pi\lambda^2=\kappa L_1 L_2$ suggesting
that the separation of the monopoles should be simply $\kappa\,L_1$
for $\kappa\leq \s012$ and $(1-\kappa)\,L_1$ for $\kappa\geq \s012$.
In fig.~\ref{history} we took $\kappa=\frac{1}{2}$ and the worldlines
are indeed separated by a half-period.  In fig.~\ref{3over8} we have
taken $\omega$ as above but increased $\kappa$ to $\frac{5}{8}$. The
first plot is the equal length case showing a peak at the centre of
the torus. At the origin of the torus, the location of the larger
constituent, there is a broader but much shallower peak. The second
plot is for $a=3$ which clearly shows two monopoles.  Moreover, the
distance between them is consistent with $\frac{3}{8}L_1$. Though the
dependence on $x_2$ is not quite as weak as for the corresponding plot
of fig.~\ref{history}; away from $\kappa=\frac{1}{2}$ more stretching
is required to access the monopole regime. \\[-15ex]

We have argued that for $a$ at or near to $1$ the constituent core
picture best describes the action density while for larger (or
smaller) $a$ almost static monopole-like objects form.  In that
respect, the instanton core and monopole constituent pictures are
complementary.  However, we have an example displaying features of
both descriptions: \\[1ex]

\begin{figure}[h]
\begin{picture}(150,150)(-250,-270)
\put(-170,-280){\epsfig{file=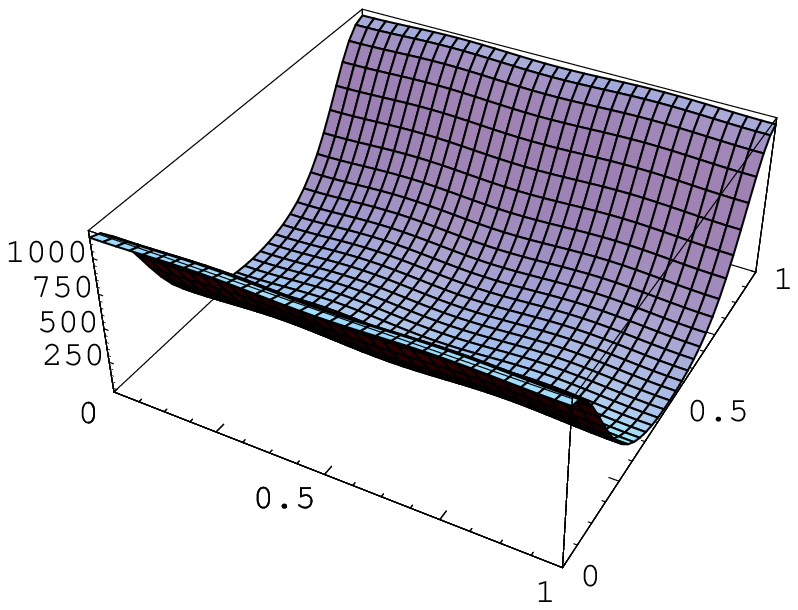,width=.50
\hsize}}
\put(-230,-180){$\bf -\s012 \Tr\,
 F^2$}
\put(-100,-265){$\bf  x_2$}  
\put(70,-230){ $\bf  x_1$}
\end{picture}
\caption{Plot of action density for $x_\perp=0$,  
$\kappa=\s012$, $a=1$ and $\omega_1=\s012 \pi, \omega_2=0$.
}
\label{monopole}
\end{figure} 

Above we plot the action density for
$\kappa=\s012$, $\omega_1=\s012 \pi$ and $\omega_2=0$. 
This configuration shows essentially no dependence on $x_2$ even
though the lengths are equal.  What seems to have occurred is that the
two cores at $x_{||}=0$ and $x_{||}=\frac{1}{2}i$ have merged into a
\it single \rm monopole. Unlike in the previous examples the
monopole worldline passes through the soft points.  That there is only
one worldline is not inconsistent with the constituent monopole
picture since by virtue of $\omega_2=0$ the `second' monopole should
be massless. The monopoles observed in figs.~\ref{history}
and \ref{3over8} have equal masses in accord with the symmetric value
of $\omega_2$.

\section{ Charge One Half}

\begin{figure}[h]
\centering
\begin{picture}(100,250)(100,0)
\epsfxsize=10cm\epsffile{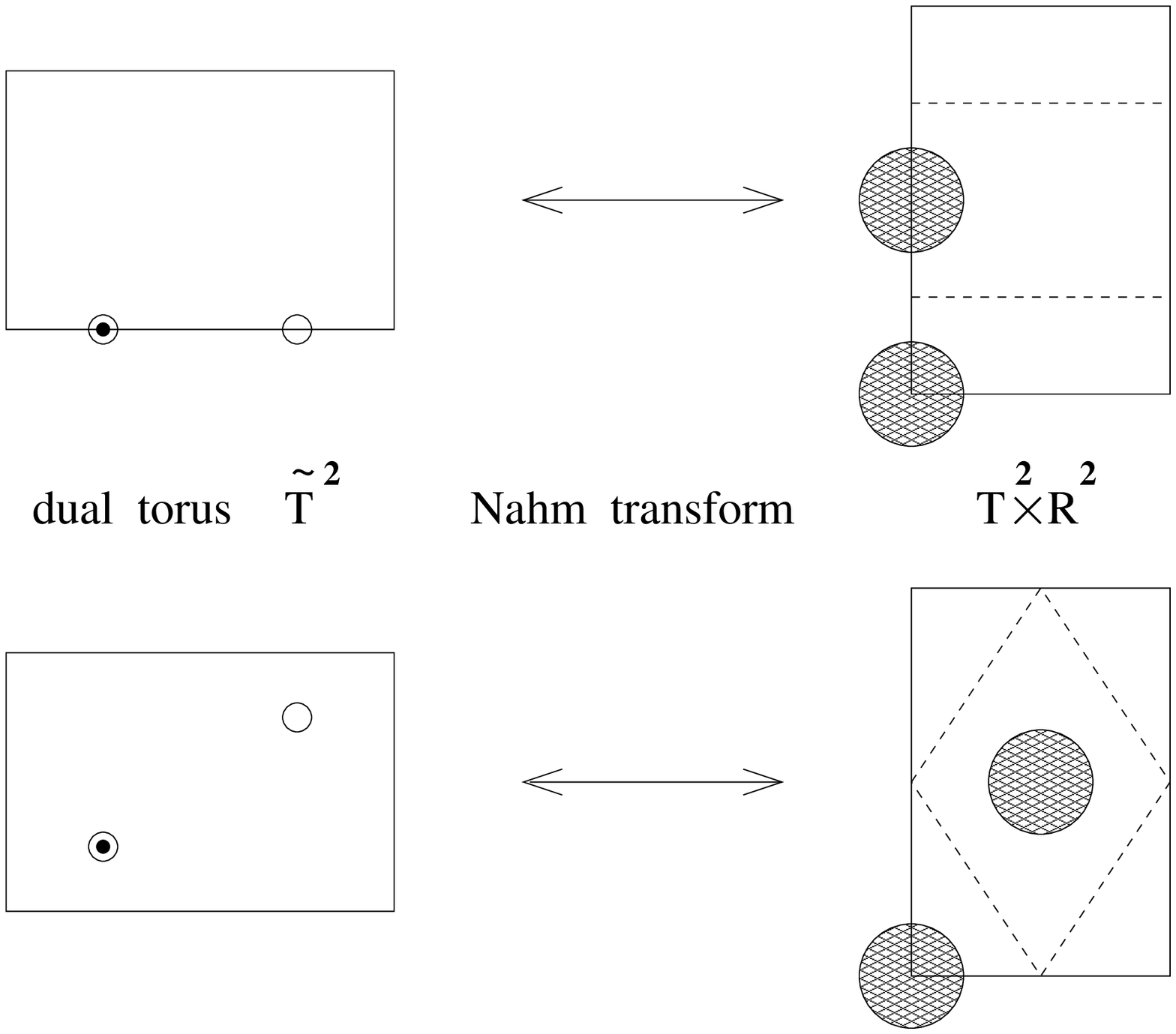}
\end{picture}
\caption{When $\kappa=\h$, the
choices $\omega_1=\h\pi/L_1$,
$\omega_2=0$ and
$\omega_1=\h\pi/L_1$,
$\omega_2=\h\pi/L_2$
yield constituent separations
allowing the torus to be cut (dashed line) to
yield charge $\h$ instantons.
}
\label{constituents}
\end{figure}

When $\kappa=\frac{1}{2}$ the two instanton cores are identical.  This
has an interesting consequence.  If we choose the constituent
locations so that they are separated by half periods the charge one
instanton can be `cut' to yield two copies of a charge $\frac{1}{2}$
instanton (see also \cite{Gonzalez-Arroyo:1998ia}). 
This happens when $(\omega_1,\omega_2)$ is
$(\frac{1}{2}\pi/L_1,0)$, $(0,\frac{1}{2}\pi/L_2)$ or
$(\frac{1}{2}\pi/L_1,\frac{1}{2}\pi/L_2)$, see
fig.~\ref{constituents}. After cutting we have a twist $Z_{12}=-\id$
in the half torus.  But to produce a half integer topological charge
 another (non-orthogonal) twist is required; this
 is most simply achieved with
$Z_{03}=-\id$.
Such a twist would have the novel feature of being associated with the
non-compact $x_0$ and $x_3$ directions. Far
away from $x_\perp=0$ the potential
 must be pure gauge
\begin{equation} 
A_\mu(x)\sim
V^{-1}(x_1,x_2,\theta)\partial_\mu V(x_1,x_2,\theta),~~~~~r\rightarrow\infty,
\end{equation}
where $x_\perp=re^{i\theta}$.
The non-compact twist translates into a double valued gauge function,
$V(x_1,x_2,\theta+2\pi)=-V(x_1,x_2,\theta)$.

The examples considered in fig.~\ref{history} are all `doubled' half
instantons.  Here the cut tori are diamond shaped (see fig. 4)
apart from the first $(a=1)$ plot, where it is  square.
 This 
case has
another interesting feature.  Compare the constituent locations with
the half instanton obtained via $L_1=\sqrt{2}$, $L_2=1/\sqrt{2}$,
$\omega_1=0$, $\omega_2=\pi/\sqrt{2}$.  On cutting both yield charge
$\frac{1}{2}$ instantons on square tori with length $1/\sqrt{2}$.  By
analogy with the four torus case, for a given set of twists and
periods, charge $\frac{1}{2}$ instantons are expected to be
unique up to
translations.  Yet these two cases have different values of $a$, one
has $a=1$ where the core picture should work while the other ($a=2$)
is expected to have some monopole characteristics.  As can clearly be
seen from the first plot in fig.~\ref{history} the action density has
maxima at the soft points.  We have computed the $a=2$ case and found
maxima {\it not} coincident with the soft points.  The soft points
correspond to saddle points of the action density.  But allowing for
translations the two half instantons are identical.  What seems to be
happening is that when $a=2$ the monopoles are still not developed,
and in this intermediate state there are lumps of action density which
while not at the soft points are identical to those at the soft points
of the $a=1$ case.  More precisely the two action densities satisfy
\begin{eqnarray}\label{relation}
\Tr\, F^2_{a=2}(x_1,x_2 )=
 \Tr\, F^2_{a=1}(\s0{1}{\sqrt{2}}(x_1+x_2)-\s014,\s0{1}{\sqrt{2}}
(x_2-x_1)-\s014). \end{eqnarray}
The $a=1$ plot is obtained by a 45 degree rotation  
of the $a=2$ plot followed by a translation. This reflects the fact that 
the two Nahm potentials are also related by a $45$ degree rotation.

\section{ Asymptotics}

In this section we consider the large $|x_\perp|$ limit of the
zero mode equations, and deduce the large $|x_\perp|$ behaviour
of the gauge potential and field strength.
This allows us to check that the charge $\frac{1}{2}$ instantons
have the expected properties.
In particular, we see that the gauge potential has asymptotic behaviour
commensurate with a  twist, $Z_{03}=-\id$.
Furthermore, the exponential decay of the action density is derived.
In fact, all the $\kappa=\frac{1}{2}$ solutions have both  the twist
and exponential decay even if they are not `doubled'
$\frac{1}{2}$ instantons.

When $|x_\perp|$ is much larger than the two periods the zero
modes of $D^\dagger_x(\hat A)$
become strongly localised about the two flux singularities.
In this regime we can approximate the zero modes with $\R^2$
solutions.
A single flux in $\R^2$
corresponds to 
$\phi=-\h\kappa \log y\bar y$.
The Weyl operator is then
\begin{equation}
-\frac{i}{2}D_x^\dagger
=\left(\begin{array}{cc}
\frac{1}{2}\bar x_\perp&
\partial_y-\frac{\kappa}{2y}-\frac{i}{2}\bar x_{||}\\
\partial_{\bar y}+\frac{\kappa}{2\bar y}-\frac{i}{2}x_{||}&
\frac{1}{2}x_\perp
\end{array}\right).
\end{equation}
When $x_\perp\neq 0$, $D_x^\dagger$ has a  normalisable
zero mode. In the special cases $\kappa=\pm \h$ the normalised modes 
 take rather simple forms:
\begin{eqnarray}
\!\!\!\psi_{\frac{1}{2}} (z)=
\frac{e^{i x\cdot z-|x_\perp|\sqrt{y\bar y}}}{\sqrt{2\pi|x_\perp|}}
\left(
\begin{array}{cr}
(y\bar y)^{-\frac{1}{4}}x_\perp\\
\frac{(y\bar y)^{\frac{1}{4}}}{\bar y}|x_\perp|
\end{array}\right)
, &\quad\!\!\!\!  & \psi_{-\frac{1}{2}}(z)=\frac{
e^{i x\cdot z-|x_\perp|\sqrt{y\bar y}}}{\sqrt{2\pi|x_\perp|}}
\left(
\begin{array}{cr}
\frac{(y\bar y)^{\frac{1}{4}}}{y}x_\perp\\
(y\bar y)^{-\frac{1}{4}}|x_\perp|
\end{array}\right),  
\end{eqnarray}
where $x\cdot z= x_1 z_1+ x_2 z_2$. For sufficiently large $|x_\perp|$
we can take
\begin{equation}\label{approx}
\psi^1(z)=\psi_{\frac{1}{2}}(z-\omega),~~~~~~~~~~~~~~~
\psi^2(z)=\psi_{-\frac{1}{2}}(z+\omega),\end{equation}
and extend the integration region from $\tilde\T^2$ to $\R^2$.
For this approximation to work 
it is important that the flux separation $2|\omega|$
is less then both dual periods, $2\pi/L_1$ and $2\pi/L_2$.
This is because we assume that the `interference'
between the fluxes dominates
the effect of the periodicity for large $|x_\perp|$. 
Note that
\begin{equation}
\int_{R^2}d^2z\,
\psi_{\frac{1}{2}}^\dagger(z-\omega)
\psi_{-\frac{1}{2}}(z+\omega)=0.\end{equation}
Inserting the zero modes (\ref{approx})
into the inverse Nahm transform (\ref{t2inverse}) produces a $U(2)$
potential; we discard the pure gauge $U(1)$ part.
The remaining $SU(2)$ piece  turns out to be in a different gauge to that
implicit in section~\ref{Greens}.
An explicit computation (see appendix C)
shows that the $SU(2)$ potential has asymptotics corresponding
to a twist and the action density  has the exponential decay
\begin{equation}\label{actiondecay}
-\frac{1}{2}\Tr F_{\mu\nu}F_{\mu\nu}\sim
\frac{32 |\omega|^3}{\pi |x_\perp|}e^{-4|\omega||x_\perp|}.
\end{equation}
We can apply this to the charge $\frac{1}{2}$ case. 
Consider such an instanton in a square box of length $L$.
This is a one instanton in the doubled box with
$L_1=L$, $L_2=2L$. Here $\omega_1=\pi/(2L)$, $\omega_2=0$,
and so the $\frac{1}{2}$ instanton has the fall-off
\begin{equation}
-\frac{1}{2}\Tr F_{\mu\nu}F_{\mu\nu}\sim
\frac{4\pi^2}{L^3|x_\perp|}e^{-2\pi|x_\perp|/L}.
\end{equation}
For a rectangular box simply replace $L$ with the longest length.
In \cite{Gonzalez-Arroyo:1998ez} a  decay exponent of $6.5$ was reported
for a unit box. This is within $5$ percent of the analytic value $2\pi$.
The decay formula
is reminiscent of similar results for Bogomolnyi vortices in the
abelian Higgs model. There the decay exponent is straightforward to obtain
 while an analytic computation of the prefactor is still an open problem
(see however \cite{tong,manton}).

One can repeat the analysis for
 $\kappa\neq \frac{1}{2}$.
The corresponding
$\R^2$ zero modes are more complicated;
they can be written in terms of modified Bessel functions.
Here the asymptotics of the gauge potential does not
correspond to a four torus twist.
More precisely,
 $\rho$ has the power decay
\begin{equation}
\rho=\frac{C}{|x_\perp|^{2\kappa}}+\hbox{
exponentially decaying remainder},
\end{equation}
where $C$ is a (dimensionful) constant.
This translates into a twist  only if $\kappa=\frac{1}{2}$.

The question of the form of the decay was an important consideration
in \cite{jardim}. There it was assumed that doubly periodic instantons
have quadratically decaying field strength components. This translates
into a quartic decay of the action density. Yet the solutions we have
 discussed show a much more rapid exponential decay.
This is not necessarily a contradiction since our discussion of the
asymptotics was confined to the radial solutions whereas \cite{jardim}
was concerned with the whole charge one sector. The  more generic non-radial
 solutions may decay quartically. Indeed the two monopole constituents
are expected to be separated in the non-compact directions. This can give
rise to an algebraic tail in the action density.

\section{Outlook}

We have developed an instanton core picture
of the simplest doubly periodic instantons.  Numerical calculations of
the action density correlate well with this picture for square tori.
However, in rectangular tori the solutions become caloron-like when
the ratio of the two periods is increased; the cores merge with
periodic copies of themselves in the short direction to form
monopole-like tubes of action density.  The basic properties of these
monopoles, such as their mass ratios and spatial separation, follow
the pattern of the charge one $SU(2)$ calorons.  It would be
interesting to develop this monopole constituent description in a more
direct manner.  One approach would be to exploit the fact that the
radially symmetric solutions considered here fall into a class of
axially-symmetric multi-calorons discussed recently
\cite{Bruckmann:2002vy} if one allows for infinite topological charge.

There are a number of obvious ways to extend the results in this
paper. An explicit treatment of the $x_\perp\neq 0$
zero mode equation is still lacking.
In the non-radially symmetric case ($\alpha\neq0$)
the zero mode equations have not yielded solutions for any
points in $\T^2\times\R^2$. Even in the absence of
explicit zero modes it might be possible
 to extract some information about possible instanton core
or monopole constituents.
A more straightforward extension would be to generalise
the results regarding radially symmetric one instantons to $SU(N)$.
That these solutions decay exponentially indicates they can be compactified
further to $\T^3\times\R$ or even $\T^4$.
The latter  is only possible for the $\kappa=\frac{1}{2}$
case where the asymptotics of the gauge field correspond to
a twist.\\[-2ex]

\noindent{\bf Acknowledgements} 

We thank F. Bruckmann, G. Dunne and P. van Baal for helpful discussions.
  C.F. was supported through a European
Community Marie Curie Fellowship (contract HPMF-CT-2000-00841).\\[.2ex]

\appendix 

\section{Properties of $f$} \label{app:proof}
Inserting the zero modes \eq{zeromodes} into the definition of $f$ 
(\ref{fdef}) yields
\begin{equation}
f(z,z';x)=\s012 (\sigma_0+i\sigma_3)e^{-\phi(z)}g_+(z,z';x)e^{-\phi(z')}+
\s012 (\sigma_0-i\sigma_3)e^{\phi(z)}g_-(z,z';x)e^{\phi(z')},
\end{equation}
where
\begin{equation}\label{gdef}
g_\pm(z,z';x)=K_\pm(z,z';x)-\frac{1}{\rho(x)}
K_\pm(z,\mp\omega;x)K_\pm(\mp\omega,z';x).
\end{equation}
It is not obvious that $f$ commutes with $\sigma_1$ and $\sigma_2$.
This only holds if the coefficients of the
projectors $\sigma_0\pm i\sigma_3$ match, that is if
\begin{equation}\label{nonquat}
e^{-\phi(z)}g_+(z,z';x)e^{-\phi(z')}
=e^{\phi(z)}g_-(z,z';x)e^{\phi(z')}.
\end{equation}
A proof of (\ref{nonquat}) for $x_\perp=0$ was given in \cite{Ford:2000zt}. 
It is sufficient to show that the left and right hand sides
of (\ref{nonquat}) have the same asymptotics at the fluxes since when
$z,z'\neq \pm \omega$ they satisfy the same
differential equation.
Consider $g_\pm(z,z';x)$ in the neighbourhood of $z=\omega$.
It follows immediately from (\ref{gdef}) that
$g_-(\omega,z';x)=0$ whereas $g_+(\omega,z';x)$ is non-zero.
This does not contradict (\ref{nonquat}) since $e^{\phi(z)}$
diverges ($\propto|y-\omega_1-i\omega_2|^{-\kappa}$) at $z=\omega$.
A short  calculation gives
\begin{equation}
g_-(z,z';x)\sim\frac{
|y-\omega_1-i\omega_2|^\kappa
}{4\pi\kappa\rho(x)}
K_-(\omega,z';x).\end{equation}
This is compatible with (\ref{nonquat})
if
\begin{equation}\label{key}
g_+(\omega,z';x)=
\frac{e^{2\phi(z')}}{4\pi\kappa\rho(x)}K_-(\omega,z';x).
\end{equation}
To check this, note that for $z'\neq\pm\omega$ both sides
satisfy the same
 differential equation and then examine them in the neighbourhoods of
$z'=\pm\omega$. 
From this discussion of the asymptotics of $g_\pm$ it should be clear
that $f(z,z';x)$, unlike
$(D_x^\dagger D_x)^{-1}(z,z')$
 is always finite. Indeed $f(z,z';x)$ tends to zero as $z$
or $z'$ approaches a flux point.

\section{Computation of \eq{fullpot}}\label{app:outline} 
Here we outline the computation of $A_{x_{||}}$; the other components
can be dealt with in much the same way. Inserting the zero modes into
the inverse Nahm transform yields
\begin{equation}
A_{x_{||}}^{11}=\frac{1}{\sqrt{\rho}}
\int_{\tilde T^2} d^2z\,
K_-(\omega,z;x)e^{\phi(z)}
\left(
D^\dagger_x(\hat A)\partial_{x_{||}} D_x(\hat A)
\right)^{11}\frac{e^{\phi(z)}}{\sqrt{\rho}}K_-(z,\omega;x).
\end{equation}
Using $(D^\dagger\partial_{x_{||}}D)^{11}=\partial_{x_{||}}(D^\dagger D)^{11}$
and the representation (33) of $(D^\dagger D)^{-1}$
\begin{equation}
A_{x_{||}}^{11}={1\over{\sqrt{\rho}}}\int d^2z
K_-(\omega,z;x)\partial_{x_{||}}\delta^2(z-\omega){1\over{\sqrt{\rho}}}=
-{1\over2}\partial_{x_{||}}\log\rho,
\end{equation}
as required. For $A_{x_{||}}^{21}$ we proceed similarly
\begin{eqnarray}
A_{x_{||}}^{21}&=&\frac{1}{\sqrt{\rho}}
\int_{\tilde T^2} d^2z\,
K_+(-\omega,z;x)e^{-\phi(z)}
\left(
D^\dagger_x(\hat A)\partial_{x_{||}} D_x(\hat A)
\right)^{21}\frac{e^{\phi(z)}}{\sqrt{\rho}}K_-(z,\omega;x)\\ \nonumber
&=& i\frac{x_\perp}{\rho}\int_{\tilde T^2} d^2z\,
K_+(-\omega,z;x)K_-(z,\omega;x).
\end{eqnarray}
The key step is to use (\ref{key})
\begin{equation}
A_{x_{||}}^{21}=4\pi i\kappa x_\perp
\int_{\tilde T^2} d^2z
K_+(-\omega,z;x)e^{-2\phi(z)}g_+(z,\omega;x).
\end{equation}
Now use (\ref{gdef}) to write $g_+$ in terms of $K_+$ and the formula
\begin{equation}
\int _{\tilde T^2}
d^2s K_+(z,s;x)e^{-2\phi(s)}K_+(s,z';x)
=-\frac{\partial}{\partial|x_\perp|^2}K_+(z,z';x),
\end{equation}
to get the result $A_{x_{||}}^{21}=-4\pi i\kappa \partial_{\bar x_\perp}
(\nu^*/\rho)$.

\section{ Exponential Decay of the Action Density}

We consider the $A_{x_\perp}$  derived from the asymptotic zero modes
given in section 9.
A straightforward calculation gives
\begin{equation}
A_{x_\perp}^{11}-A_{x_\perp}^{22}=0.\end{equation}
The off-diagonal components are more involved:
\begin{equation}
A_{x_\perp}^{21}=
\int_{R^2}d^2z\,
\psi_{-\frac{1}{2}}^\dagger(z+\omega)\frac{\partial}{\partial x_\perp}
\psi_{\frac{1}{2}}(z-\omega)=\frac{\bar x_\perp}{2\pi |x_\perp|}
e^{2ix\cdot\omega} (I+J).
\end{equation}
Here
\begin{equation}
\label{Iint}
I=
\int_{R^2} d^2z\frac{(y_2\bar y_2)^{\frac{1}{4}}}
{\bar y_2}(y_1\bar y_1)^{-\frac{1}{4}}
e^{-|x_\perp|(\sqrt{y_1\bar y_1}+\sqrt{y_2\bar y_2})}
\end{equation}
and 
\begin{eqnarray}\label{Jint}
\lefteqn{J = {1\over2}
\int_{R^2}
 d^2z e^{-|x_\perp|(\sqrt{y_1\bar y_1}+\sqrt{y_2\bar y_2})}}\hspace{2cm}
\\\nonumber
& & \times\left[
(y_2\bar y_2)^{-{1\over4}}{(y_1\bar y_1)^{{1\over4}}\over{
\bar y_1}}-|x_\perp|{(y_2\bar y_2)^{{1\over4}}\over{
\bar y_2}}(y_1\bar y_1)^{{1\over4}}-
|x_\perp|(y_2\bar y_2)^{-{1\over4}}
{(y_1\bar y_1)^{{3\over4}}\over{
\bar y_1}}\right],
\end{eqnarray}
where $y_1=y-\omega_1-i\omega_2$ and $y_2=y+\omega_1+i\omega_2$. 
As $|x_\perp|$ is increased the integral $I$ is
dominated by a small neighbourhood about 
the line-segment joining the two fluxes.
For example, if $\omega_2=0$ we have
$e^{-|x_\perp|(\sqrt{y_1\bar y_1}+\sqrt{y_2\bar y_2})}
\sim\pi^{\h}|x_\perp|^{-\h}|\omega_1|^{-\h}
(\omega_1^2-z_1^2)^{\h}
e^{-2|x_\perp||\omega_1|}\delta(z_1)$
for $-\omega_1<z_1<\omega_1$.
Integrating along the line-segment gives
\begin{equation}
I\sim
\frac{2\pi^{\h}|\omega|^{-\h}|x_\perp|^{-\h}}{
\omega_1-i\omega_2}
e^{-2|\omega||x_\perp|}
.
\end{equation}
$J$ exhibits a similar exponential decay. 
$A_{\bar x_\perp}^{21}$ is simply $x_\perp e^{2ix\cdot\omega}J/
(2\pi |x_\perp|)$, and so
\begin{equation}
F_{x_\perp \bar x_\perp}^{21}=- 
\frac{1}{2\pi |x_\perp|}
e^{2ix\cdot\omega}\partial_{\bar x_\perp}(\bar x_\perp I)
\sim e^{2ix\cdot \omega}
e^{-2|\omega||x_\perp|}
\frac{(\omega_1+i\omega_2)|\omega|^{\h}}
{\pi^{\h}|x_\perp|^{\h}}
\end{equation}
We also have
\begin{equation}
F_{x_\perp \bar x_\perp}^{11}=O(e^{-4|\omega||x_\perp|}).
\end{equation}
Now compare with the corresponding field strengths generated by the ansatz
(\ref{fullpot}) in section~\ref{Greens}.
\begin{equation}
F_{x_\perp\bar x_\perp}^{21}=2\pi i\rho\partial_{\bar x_{||}}\partial_{
\bar x_\perp}
\frac{\nu^*}{\rho},~~~~~~~~~~~~
F_{x_\perp\bar x_\perp}^{11}=\partial_{x_\perp}
\partial_{ \bar x_\perp}\log\rho
+(2\pi\rho)^2\partial_{x_{||}}\frac{\nu}{\rho}\partial_{\bar x_{||}}
\frac{\nu^*}{\rho}.\end{equation}
Consider
\begin{equation}\label{fit}
\rho=\frac{C}{|x_\perp|^t},~~~~~~~~~~
\nu^*=\frac{1}{2}\pi^{-\frac{3}{2}}
e^{2i\omega\cdot x} e^{-2|\omega||x_\perp|}|\omega|^{-\frac{1}{2}}
|x_\perp|^{-\frac{1}{2}},\end{equation}
where $t$ is some real number and $C$ is a  constant. 
Neglecting sub-leading terms the expressions for $F_{x_\perp\bar
x_\perp}$ agree up to a phase,
$\sqrt{x_\perp/\bar x_\perp}=e^{i\theta}$,
which can be compensated for by the double-valued gauge 
transformation $V=e^{{1\over2}i\theta\tau_3}$. Under this gauge transformation
 the diagonal part of $A_{x_\perp}$ becomes non-trivial
\begin{equation}
A_{x_\perp}=\frac{\tau_3}{4x_\perp}+
O(e^{-4|\omega||x_\perp|}).
\end{equation}
which agrees with \eq{fullpot} if  $t=1$ in (\ref{fit}).
This is exactly what is required to generate the twist $Z_{03}=-\id$.
Using the large $|x_\perp|$ form of $\rho$ and $\nu$
it is straightforward to obtain the decay formula
(\ref{actiondecay}).

\end{document}